\documentclass[a4paper,fleqn,usenatbib]{mnras}
\usepackage{mathptmx}

\usepackage[T1]{fontenc}
\usepackage{ae,aecompl}

\usepackage{graphicx}	
\usepackage{amsmath}	
\usepackage{amssymb}	
\usepackage{color}

{\newcommand{\gcm}{\, {\rm g\, cm}^{-3}}

\def\simlt{\mathrel{\hbox{\rlap{\hbox{\lower4pt\hbox{$\sim$}}}\hbox{$<$}}}}
\def\simgt{\mathrel{\hbox{\rlap{\hbox{\lower4pt\hbox{$\sim$}}}\hbox{$>$}}}}

\title[Submoons]{Can Moons Have Moons?}

\author[Kollmeier \& Raymond]{
Juna A. Kollmeier$^{1}$\thanks{E-mail: jak@carnegiescience.edu} \&
Sean N. Raymond$^{2}$\thanks{E-mail: rayray.sean@gmail.com}\\
$^{1}$ Observatories of the Carnegie Institution of Washington, 813 Santa Barbara St., Pasadena, CA 91101\\
$^{2}$ Laboratoire d'Astrophysique de Bordeaux, Univ. Bordeaux, CNRS, B18N, all{\'e} Geoffroy Saint-Hilaire, 33615 Pessac, France\\
}

\date{Accepted XXX. Received YYY; in original form ZZZ}

\pubyear{2018}

\begin{document}
\label{firstpage}
\pagerange{\pageref{firstpage}--\pageref{lastpage}}
\maketitle

\begin{abstract}

Each of the giant planets within the Solar System has large moons but none of these moons have their own moons (which we call {\em submoons}). By analogy with studies of moons around short-period exoplanets, we investigate the tidal-dynamical stability of submoons.  We find that 10 km-scale submoons can only survive around large (1000 km-scale) moons on wide-separation orbits. Tidal dissipation destabilizes the orbits of submoons around moons that are small or too close to their host planet; this is the case for most of the Solar System's moons. A handful of known moons are, however, capable of hosting long-lived submoons: Saturn's moons Titan and Iapetus, Jupiter's moon Callisto, and Earth's Moon. Based on its inferred mass and orbital separation, the newly-discovered exomoon candidate Kepler-1625b-I can in principle host a large submoon, although its stability depends on a number of unknown parameters.  We discuss the possible habitability of submoons and the potential for subsubmoons.  The existence, or lack thereof, of submoons, may yield important constraints on satellite formation and evolution in planetary systems.  
\end{abstract}



\section{Introduction}
In all known planetary systems, natural satellites occur in a restricted dynamical phase space: planets orbit stars and moons orbit planets.  It is natural to ask, can moons have their own stable satellites (submoons)?  If so, why don't any of the known moons of the Solar System have their own submoons?  One possibility is that the formation mechanism of planet-moon systems precludes their formation.  Another possibility, is that these bodies are dynamically unstable and are rapidly scoured from their system after formation.  Here, we investigate the latter hypothesis.

What are the requirements for stability of a submoon? To ensure dynamical stability, the host moon must have a Hill sphere that is larger than its physical radius as well as its Roche limit. The submoon must also survive any long-term dynamical effects such as tidal evolution.

Tidal stresses deform extended objects and internal dissipation leads to changes in the objects' rotation states and orbits~\citep[e.g.,][]{darwin1879,goldreich66,ferrazmello08}. Tidal evolution in a planet-moon system generally causes the moon's orbit to widen if the planet spins quickly or to shrink if the planet spins slowly~\citep[e.g.][]{peale80,burns86}.

Tidal evolution in star-planet-moon systems has been studied in the context of Venus and Mercury's lack of moons~\citep{counselman73,ward73,burns73} and in the more general case of moons orbiting exoplanets on short-period orbits~\citep{barnes02,sasaki12,sasaki14,piro18}. In a star-planet-moon system, stellar tidal friction acts to slow the planet's rotation, with a direct consequence for the moons' tidal migration~\citep{ward73,burns73}. Depending on the configuration, moons may migrate inward and crash into their host planets or migrate outward until they reach the stability limit. In some cases moons can first migrate outward, then change direction and migrate inward as the planet spins down~\citep{barnes02,sasaki12,piro18}. 

Here we apply this concept to planet-moon-submoon systems. \cite{barnes02} showed that under tidal evolution there is a maximum mass of a moon that can survive for a given time $T$ around a close-in exoplanet. To allow for a comparison with the known moons, we re-frame their analysis to ask: {\em what are the physical and orbital requirements for a moon to host a stable submoon with specified properties?}  

Our paper is structured as follows. We present simple calculations of submoon survival in \S 2.  We then discuss the long-term stability of submoons (\S 3) as well as their potential habitability (\S 4).  We conclude in \S 5, and discuss why subsubmoons -- moons orbiting submoons -- can only exist at very small sizes.

\section{Tidal Calculations}
Our goal is to map the parameter space in which submoons may be stable under the action of planet-moon-submoon tides. We follow \cite{barnes02}, who derived a simple analytical approach that applies across different outcomes of tidally-driven migration. We adapt Eq. 8 of \cite{barnes02} to derive the critical size of a moon $R_{moon}$ that can host a long-lived submoon.  We find that
\begin{equation}
    R_{moon} \ge \left[ \frac{39 M_{sub} \, k_{2,moon} \, T \sqrt{G}}{2 \left(4 \pi \rho_{moon}\right)^{8/3} Q_{moon}} \left(\frac{3 M_p}{\left(f \, a_{moon}\right)^3}\right)^{13/6}\right]^{1/3},
    \label{eqn:stability}
\end{equation}

\noindent where $M_{sub}$ is the (fixed) mass of the submoon in question, $R_{moon}$, $a_{moon}$, $\rho_{moon}$, $Q_{moon}$  and $k_{2,moon}$ are the moon's radius, orbital radius, bulk density, tidal quality factor, and tidal Love number, respectively, $M_p$ is the planet mass, $T$ is fixed at 4.6 Gyr, and $G$ is the gravitational constant. A submoon's orbit is stable out to a fraction $f$ of its moon's Hill sphere; \cite{domingos06} showed that $f \approx 0.4895$ for prograde, low-eccentricity orbits. Eq. 1 has the advantage of simplicity while capturing the basic mechanisms at play.  However, we note that it inherently assumes that the submoon is low-mass and neglects effects such as the influence of the submoon on the moon's rotation, the influence of the moon on the planet's rotation and the effect of the initial moon and planet rotation; see \cite{sasaki12} and \cite{piro18} for a more comprehensive treatment.

\begin{figure*}
\centering
\includegraphics[scale=0.95]{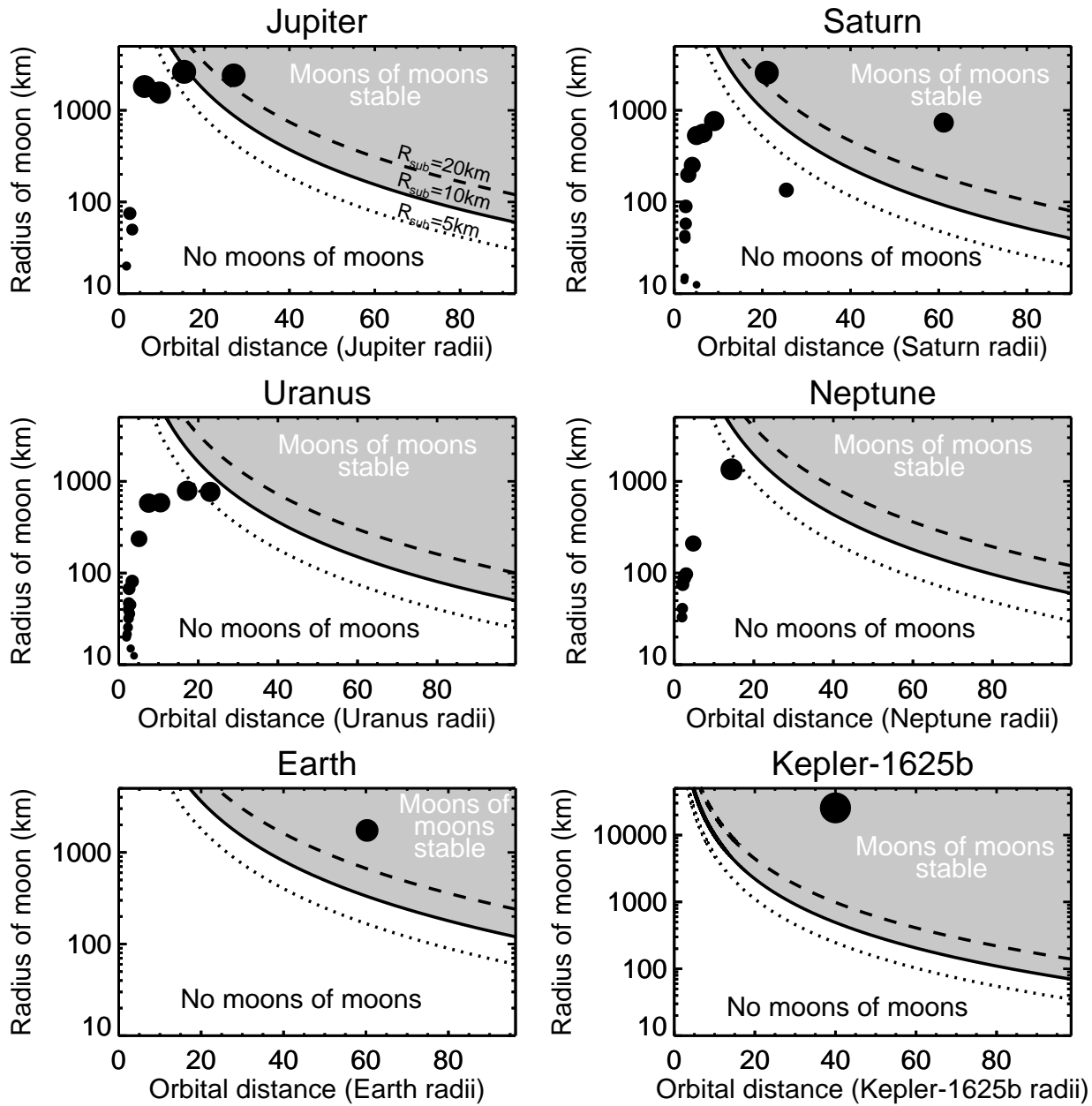}
\caption{{\bf Moons of Moons} -- The parameter space in which the moon of a specified planet could host a long-lived submoon under the action of planet-moon-submoon tides. A submoon would be stable for at least the age of the Solar System in the shaded region to the upper right of each panel. The default size of the submoon is 10km in radius (solid curves), and we also show the critical limits for submoons of 5km (dotted) and 20km (dashed) in radius. The solid dots are each planet’s actual satellites.  This calculation assumes that all moons have bulk densities of 2.5$\gcm$, appropriate for most large satellites of the giant planets. We assume tidal Love numbers k$_{2,moon} = $0.25~\citep{moore00}, and tidal quality factor Q = 100~\citep{lainey16}. We assumed that all submoons have densities of 2$\gcm$, generally appropriate for smaller bodies. For the case of the Earth we assumed a moon density of the Lunar density of 3.34$\gcm$. For the case of Kepler-1625b, we assumed a planet mass of 4M$_J$~\citep{teachey18}, k$_{2,moon} = $0.12~\citep[appropriate for the ice giants;][]{gavrilov77} and a very uncertain $Q_{moon}$ of 1000~\citep[see][]{lainey16}.}
\label{fig:submoons}
\end{figure*}

Figure~\ref{fig:submoons} shows the regions of parameter space where a long-lived, 10~km-scale submoon could exist under the action of planet-moon-submoon tides (see caption for parameter choices). Only large moons on wide-separation orbits can host long-lived submoons. This is mainly because massive, distant moons have larger Hill radii that provide more stable volume for submoons. Most of the large regular moons of the giant planets are too close to their host planets to host submoons. This is the case for all of Uranus and Neptune's moons. If submoons did form in these systems, they have since been removed by tidally-induced migration. 

\begin{figure*}
\centering
\includegraphics[scale=0.45]{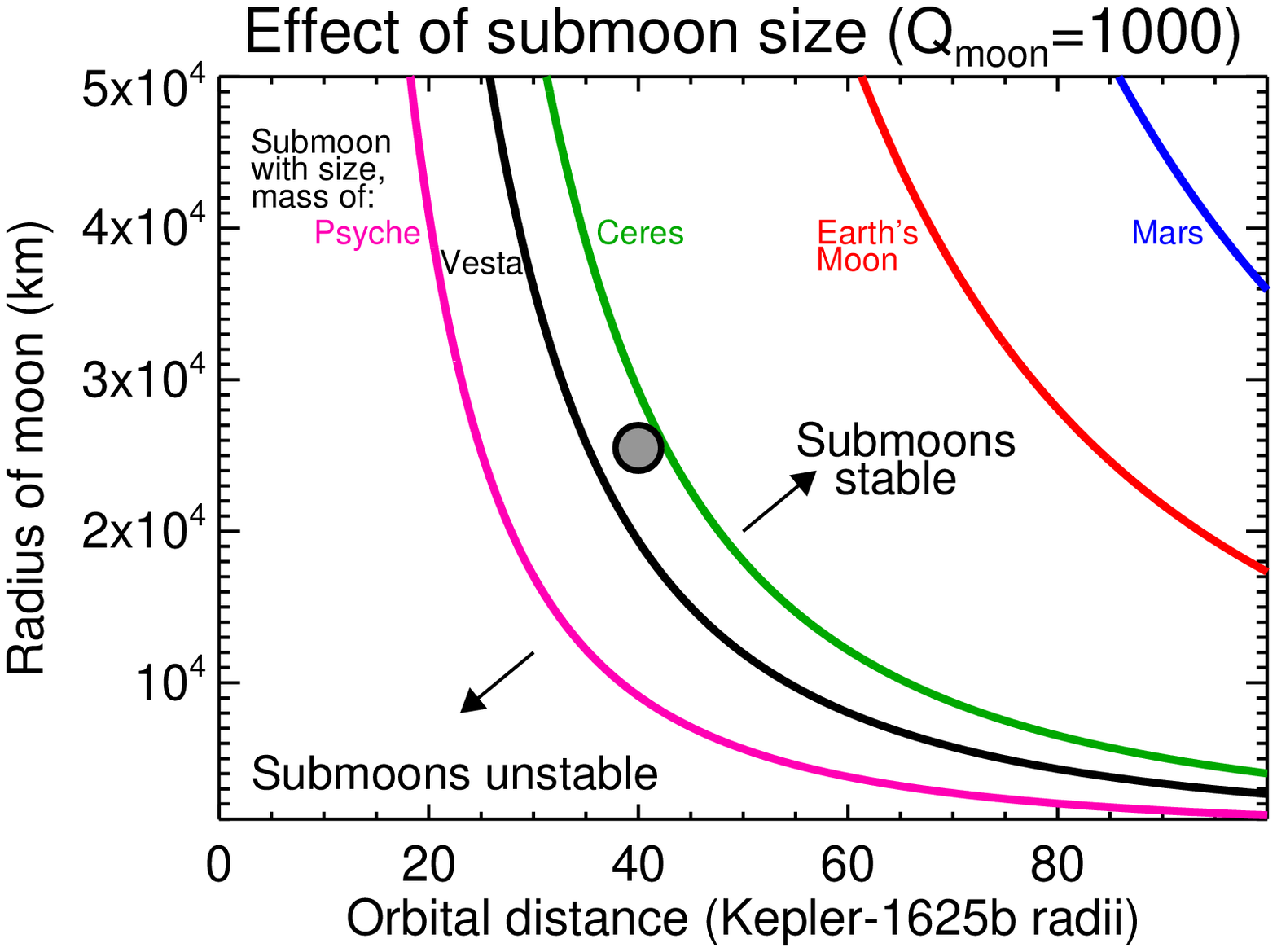}
\includegraphics[scale=0.45]{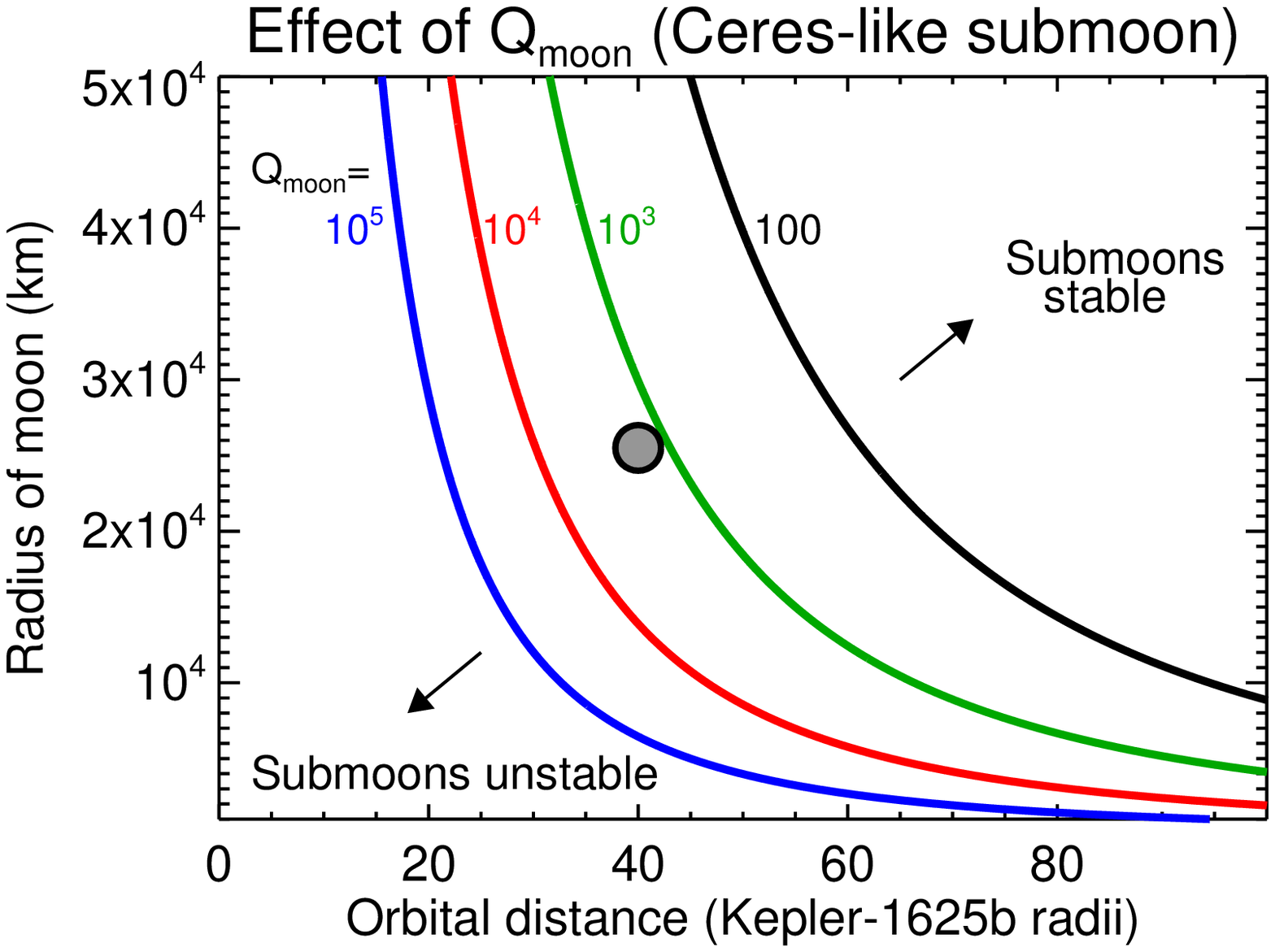}
\caption{Effect of the submoon size (left) and moon tidal quality factor $Q_{moon}$ (right) on the ability of a moon to host a long-lived submoon. Submoons are long-term stable above and to the right of each curve.  We assume $M_p=4M_J$~\citep{teachey18}, k$_{2,moon} = $0.12~\citep{gavrilov77}, and $T$ = 9 Gyr~\citep{teachey18}. In the left hand panel $Q_{moon}$ is fixed at 1000 and in the right hand panel the submoon is fixed at Ceres' size and mass.}
\label{fig:params}
\end{figure*}

Remarkably, Jupiter (Callisto), Saturn (Titan and Iapetus), and Earth (Moon) each have the potential to host long-lived submoons around their current moons. Based on its orbital separation and inferred mass and size, the new exomoon candidate Kepler-1625b-I~\citep{teachey18} also appears capable of hosting a large submoon. However, it is worth noting that Kepler-1625b-I has a significant orbital inclination which may affect the stability of submoons~\citep[e.g.][]{tremaine09,tremaine14,grishin17}. We encourage detailed studies of the dynamical stability of submoons in this system in particular, but also in the general case of inclined orbits where additional dynamical effects may play a role in long-term submoon stability.

We use the Kepler-1625b system to illustrate the effect of relevant system parameters.  Assuming the orbit of its candidate moon to be coplanar, Figure~\ref{fig:params} shows the effect of the submoon size/mass and the tidal quality factor $Q_{moon}$ on the parameter space available for long-lived submoons for an adopted system age of 9 Gyr~\citep[]{teachey18}. Smaller submoons are stable over a broader range of moon radius and distance from their host planet than larger submoons.  As expected from Equation~\ref{eqn:stability}, the ``survivability space" for submoons increases with increasing $Q_{moon}$, and for decreasing submoon mass and density. At face value the exomoon candidate Kepler-1625b-I could in principle host a Vesta- to Ceres-like submoon. This is consistent with the results of \cite{reid73}, who found that a factor of $\sim 10^{-5}$ in mass serves as a rough guideline for the potential stability of a long-lived submoon. Of course, it is not clear that such a submoon could survive in the face of other dynamical effects. Indeed, no such submoons have survived in the Solar System, which leaves the open question:  did they form and get removed by other effects than we have considered here, or did they never form at all?

\section{Long-term survival of submoons}
Our calculations show that submoons may remain stable to tidal evolution in orbit around Callisto, Iapetus and the Moon. So why don't they those moons host submoons? 


The tidal-dynamical stability described in \S 2 is only one criterion for long-term survival of a submoon. Of course, for submoons to exist they must have a formation pathway.  The large moons of the gas giants are thought to have formed in circum-planetary disks~\citep[e.g.][]{canup06,cilibrasi18} or by spreading of dense primordial ring systems~\citep{charnoz10,crida12}. Earth's large moon is thought to have formed via a giant impact~\citep[e.g.][]{benz86,canup04}, whereas Mars' small moons may have been captured or created after a large impact~\citep[see][]{rosenblatt11}. 

Even within a moon's presumably stable region there are other sources of dynamical instability that may remove submoons on certain orbits. For example, the Moon has been found to have localized mass concentrations within its crust~\citep{muller68} that destabilize the orbits of very close orbits around the Moon~\citep{konopliv01}. More broadly, a range of orbits around the Moon are destabilized by perturbations from the Sun and Earth~\citep{lidov63,scheeres98,elipe03}. Such perturbations are likely to play a role in destabilizing the orbits of submoons such that the available parameter space for submoon stability is more restricted than what we derived in \S 2.  Detailed calculations of this are beyond the scope of this work, but we note further work must be done to investigate these effects in any realistic system.

If primordial moons did form around Callisto, Iapetus or the Moon they must later have been removed.  One mechanism for submoon removal is its host moon's tidally induced migration, during which a submoon's orbit may shrink or grow~\citep{namouni10} and can become trapped in an unstable evection resonance~\citep{spalding16}. In addition, if a moon underwent significant outward migration then its ability to host a primordial submoon depends on its initial orbital radius rather than its final one.  For instance, the Moon is thought to have migrated outward from just a few Earth radii~\citep[e.g][]{goldreich66b,touma94}.  Yet submoons of the Moon would only have been stable once the moon was beyond $\sim 30$ Earth radii.  Thus, the Moon could only host a very young submoon.

Iapetus's equatorial ridge may provide evidence for a past submoon.  \cite{levison11} and \cite{dombard12} proposed that a collision produced a submoon orbiting Iapetus.  \cite{levison11} argued that the submoon-generating collision also produced a closer belt of debris. The submoon tidally evolved outward whereas the debris was tidally pushed inward to create Iapetus' ridge.  In contrast, \cite{dombard12} proposed that the ridge was created by the tidal in-spiralling and shredding of the submoon itself.  It has also been proposed that primordial submoons of Earth's moon were destabilized and crashed into the Moon~\citep{reid73,conway86}. 
 
Dynamical interactions between moons could also have a destabilizing effect on submoons.  Secular perturbations in compact planetary systems decrease the stability radius for moons, effectively decreasing the critical distance for satellite stability to $f\approx 0.4$~\citep{payne13}.  By analogy, submoons could have less stable volume in multiple-moon systems, although we do not expect these effects to dominate, particularly at late-times.  In addition, dynamical scattering events between planets often destabilize the orbits of host moons~\citep{gong13,hong18,rabago18}.  Moon-moon scattering events may be common in satellite systems~\citep{perets14}, particularly at early times, and may serve to destabilize submoons.  Further calculations beyond the scope of this work are required to determine the amplitude of these effects for different primordial scenarios.

\section{Potential habitability of submoons and other speculative scenarios}
Moons of gas giant planets have long been considered potential sites for extant life~\citep{williams97,scharf06,heller14, forgan18}.  One may then wonder: {\em can submoons be habitable?}

If plate tectonics and a long-lived atmosphere are prerequisites for habitability, then there may be a lower limit on the mass of potentially habitable world. \cite{williams97} derived a rough lower limit of $0.23 M_\oplus$~\citep[see also][]{raymond07}. At twice the mass of Mars, this is far more massive than any submoons that appear to be stable around any Solar System moons or the Kepler-1625b-I exomoon candidate (see Figs. 1 and 2).  

How can we imagine a habitable system in which a large submoon is stable to tides? From Eq. 1 we can see that the mass of the maximum stable submoon $M_{sub, max} \propto M_{moon} \, Q_{moon} \, \left(a_{moon}^3/M_p\right)^{13/6}$. This is sensible: submoons should survive longer when tides are weak.  Given the very strong scaling of tidal evolution with separation~\citep[e.g.][]{goldreich66,ferrazmello08}, this implies that large-scale systems are the most likely to host large submoons. Submoons also are more likely to exist around moons on wide orbits around their planets that have large Hill spheres.  Of course, the widest possible moon orbit is linked with a planet's Hill sphere such that, if we invoke a moon that orbits at a fixed fraction of its parent planet's Hill sphere, then $M_{sub, max} \propto \left(a_p^3/M_\star\right)^{13/6}$, where $M_\star$ is the stellar mass.  If we require the planet to be located in its host star's circumstellar habitable zone, scaled simply as $a_{HZ} = 1 {\rm AU} \, \sqrt{L/L_{\odot}}$~\citep{kasting93,selsis07,kopparapu13}, then can use the known stellar mass-luminosity relationship $(L_\star/L_\odot) \propto (M_\star/M_\odot)^{\alpha}$, where $\alpha$ ranges from 2 to 4 for the main sequence stars of interest, to set the boundaries within which the submoon can be both dynamically long-lived and habitable:

\begin{eqnarray}
    M_{sub, max, HZ} &\propto& \left(a_{HZ}^3/M_\star\right)^{13/6}\\
    &\propto&  M_\star^{\left( \frac{13\alpha}{4} - \frac{13}{6}\right)}.
\label{hab}
\end{eqnarray}

This implies that the maximum submoon mass is a very strong function of the stellar mass (e.g., for $\alpha = 4$, $M_{sub,max,HZ} \propto M_\star^{65/6}$!). This is a consequence of the strong scaling of tidal effects, compounded with the physical size of satellite systems. For more massive stars, the habitable zone is more distant and habitable zone planets have wider Hill spheres. Large moons can therefore survive on more distant orbits and have larger Hill spheres such that submoons can exist on wider orbits where submoon-moon-planet tidal evolution is far slower, as compared with lower-mass stars.  

The arguments above only apply if moons' orbital radii scale with their host planets' Hill radii.  If, on the other hand, the orbital separations of moons is governed by a different process such that moons' orbital radii are independent of their planets' Hill radii, then the maximal submoon mass would be more or less independent of the stellar mass.

Our study also opens the tantalizing subject of the largest object that could be built (presumably by humans) around the Earth's moon and remain stable: a non-natural submoon.  Such an outpost may be long-lived, provided its orbit is dynamically stable, and reasonably sized.  We can calculate the maximum potential size of our non-natural submoon in a similar manner as above, although information about the material properties of such a structure would be required.  Similarly, more sophisticated analysis would be required to determine the viability of such a structure from a materials point of view.  

\section{Conclusions}
We have shown that, in the face of tidal evolution, submoons can only survive around massive moons on wide orbits around their host planets.  We acknowledge the pioneering paper of \cite{reid73}, who was the first, to our knowledge, to study the stability of submoons under the effect of tides, and who derived an approximate mass ratio of $10^{-5}$ for the mass of a long-lived submoon. Our calculations are consistent with those results.

Given that some moons do appear capable of hosting submoons, one may further wonder whether submoons could host their own satellites, or {\em subsubmoons}.  Using \cite{reid73}'s rule of thumb, a subsubmoon would have to be less than roughly $10^{-5}$ times the submoon's mass.  For the case of the exomoon candidate Kepler-1625b-I~\citep{teachey18} we showed in Fig.~\ref{fig:params} that the largest stable submoon is Vesta- to Ceres-sized.  The largest possible submoon would thus be roughly 5-10 km in radius.  For much less massive, Solar System-like moons the largest stable submoon was $\sim 10$~km such that the largest possible subsubmoon would be sub-km-sized.

To conclude, we note that while many planet-moon systems are not dynamically able to host long-lived submoons, the absence of submoons around known moons and exomoons where submoons {\it can} survive provides important clues to the formation mechanisms and histories of these systems.  Further studies of the potential formation mechanisms, long-term dynamical survival, and detectability of submoons is encouraged.

\section*{Acknowledgements}
We thank the referee Jason Barnes for a constructive review. We are grateful to Scott Tremaine, Scott Kenyon, Caleb Scharf, Michele Bannister and Dan Fabrycky for useful discussions.  We are particularly grateful to Levi J. Buckwalter, whose initial query from 2014 inspired this work. S.~N.~R. thanks the Agence Nationale pour la Recherche for support via grant ANR-13-BS05-0003-002 (grant MOJO) and acknowledges NASA Astrobiology Institute’s Virtual Planetary Laboratory Lead Team, funded via the NAI under solicitation NNH12ZDA002C and cooperative agreement No. NNA13AA93A. This research was supported in part by the National Science Foundation under Grant No. NSF PHY11-25915.




\bibliographystyle{mnras}
\bibliography{refs_moonmoons} 




\end{document}